 \documentclass[prd,amsmath,amsfonts,11pt,superscriptaddress,onecolumn,nofootinbib]{revtex4}
\usepackage{graphicx}
\usepackage{dcolumn}
\usepackage{bm}
\usepackage{amsfonts}

\usepackage[dvips]{color}

\usepackage{mathrsfs}
\usepackage{amsmath}
\usepackage{amsfonts}
\usepackage{amssymb}

\usepackage{graphics}
\usepackage{amsmath}
\usepackage{amsfonts}
\usepackage{amssymb}
\usepackage{graphicx}
\usepackage{epsfig}

\def\ba{\begin{eqnarray}}
\def\ea{\end{eqnarray}}
\def\be{\begin{equation}}
\def\ee{\end{equation}}

\def\nn{\nonumber}

\def\d{\mathrm{d}}

\def\mn{_{\mu \nu}}
\def\mupn{^\mu_{\, \nu}}
\def\({\left(}
\def\){\right)}

\def\kf{\kappa_5^2}
\def\kq{\kappa_4^2}

\def\ie{{\it i.e.} }
\def\bv{\bar v}

\begin{document}

\title{Living on a dS brane: Effects of KK modes on inflation}
\author{Claudia de Rham}
\email{crham@perimeterinstitute.ca}
\affiliation{Dept. of Physics \& Astronomy, McMaster University, Hamilton ON, Canada}
\affiliation{Perimeter Institute for Theoretical Physics,  Waterloo, ON, Canada}
\author{Scott Watson}
\email{watsongs@physics.utoronto.ca}
\affiliation{Physics Department, University of Toronto, Toronto, ON, Canada}
\date{\today}

\begin{abstract}
We develop a formalism to study non-local higher-dimensional effects in
braneworld scenarios from a four-dimensional effective theory point
of view and check it against the well-known Garriga-Tanaka result in the appropriate limit.
We then use this formalism to study the spectrum of density
perturbations during inflation as seen from the lower-dimensional effective theory.
In particular, we find that the gravitational potential is greatly enhanced
at short wavelengths. The consequences to the curvature
perturbations are nonetheless very weak and will lead to no
characteristic signatures on the power spectrum.
\end{abstract}
 \maketitle
\nopagebreak

\section{Introduction}
As an alternative to Planck-scale string compactifications,
braneworld models have recently received an increased interest from both
cosmologists and phenomenologists.
One reason is because such theories generically predict consequences of quantum theories of gravity
at an energy scale much lower than one would naively expect.
As a consequence, this opens the possibility of probing for experimental signatures coming from IR/UV modifications to the low-energy effective theory
at upcoming experiments.
Moreover, this also allows for model building, which can then attempt to unravel outstanding puzzles in both
particle phenomenology (e.g. hierarchy problems), and cosmology (e.g. the origin and nature of inflation and dark
energy) by exploring consequences of the fundamental theory.

Within these models, the presence of extra-dimensions and
the associated non-local terms are of particular interest.
A common approach to treating such models is to utilize the
low-energy effective theory, which is expected to be valid at scales comparable or less than the
Hubble scale today.  In the context of the Randall-Sundrum model
\cite{Randall:1999vf}, this approach has been used extensively and the
low-energy effective theory, first developed in
\cite{lowenergy}, gives rise to a scalar-tensor
theory of gravity.

At higher energies, however, such as those occurring in the early
universe, the gravitational behavior of braneworld scenarios is significantly different from
any conventional scalar-tensor cosmology.
Understanding the behavior of these models during, for instance,
inflation can then provide an opportunity to find
characteristic signatures of such higher-dimensional models.
Motivated by this possibility, effective theories including
characteristic high-energy corrections have been
developed in \cite{Barvinsky:2002kh,deRham:2004yt}.
In these models, corrections to the effective theory were considered arising from higher dimensional
Kaluza-Klein (KK) modes in a Minkowski background.
Although this approach gives an important insight into the
higher-dimensional effects, it is more desirable to consider
a brane which more closely resembles the observed universe, namely one of the
Friedmann-Robertson-Walker type. In this paper, we will develop
such an effective theory, allowing us to
understand the effect of the entire non-local Kaluza-Klein (KK) continuum
at the perturbed level around a single de Sitter (dS)/inflationary brane.

One common approach to brane inflation model building is to consider the inflaton scalar responsible for inflation
to result from one of the many moduli of the theory, such as the size or shape of the extra dimensions, or the
relative position or orientation of the branes\footnote{See e.g. \cite{Cline:2006hu} and references within.}.
However in this study, we take an alternative perspective and
explore the possibility of having an inflaton scalar field confined to the
brane. This approach has been consider previously in
\cite{braneinflation}, where it was argued that the usual constraints on slow-roll inflation models (e.g. slope of the potential)
may be partially relaxed due to the modifications of the effective gravity theory from the presence of extra dimensions.

Here we want to extend the analysis performed in \cite{braneinflation} to incorporate the
effect of the entire KK continuum, which is often neglected in the
literature. To this end, we first recall
how the KK tower of \cite{Garriga:1999yh} can be derived using the
Gauss-Codacci formalism for the Minkowski brane as shown in \cite{Shiromizu:1999wj}. We then extend this analysis to
the case of an inflationary brane and argue that within the slow-roll regime
this theory remains a successful model of inflation
even in the presence of the entire non-local KK tower.
This result is certainly not surprising given the anticipated scaling of the corrections
to the low-energy effective theory.  In fact, we will find this scaling explicitly by finding the influence of the KK tower
on the initial state of the scalar perturbations.
In particular, we find that the gravitational potential is
greatly enhanced at short wavelengths, but explain why this enhancement is
not significant enough to produce any noticeable signature during
inflation (to leading order in the slow-roll parameters.)
We will also  discuss how our approach can be used to study
the spectrum of gravity waves  and the effect of higher dimensional
anisotropy on the lower dimensional effective theory.

The paper is organized as follows.
In Section \ref{section effective theory on dS brane}, we present the effective theory for an inflationary brane embedded in an AdS bulk
and discuss the main departures from the standard low-energy effective theory.
In the next section, we consider as an application the behavior of scalar metric perturbations during the modified gravity inflation.
We will consider the short-wavelength modification to both dynamics and initial conditions, showing
that the resulting correction to the power spectrum is suppressed.  This is our main result.
We then conclude in Section \ref{section conclusion}.
In the appendices we provide a more detailed derivation of the
effective theory resulting from the inclusion of the KK tower.
We first consider the Minkowski case, followed by the case of a dS/inflationary brane.
We then compare both
results in the Minkowski limit and check the validity of our theory
in Appendix \ref{appendix check}. The short-wavelength limit of
this theory is then developed in Appendix \ref{appendix short} for
the special case of scalar perturbations during inflation.

\section{KK tower on a Inflationary brane}
\label{section effective theory on dS brane}

In this section we first develop a new formalism to make the problem
of finding the low-energy effective theory more technically tractable.
We will
then use the new formalism to study perturbations around an
inflating brane.

\subsection{Formalism}
We want to consider the one-brane Randall-Sundrum II model \cite{Randall:1999vf}, with a positive tension $D3$
brane localized at the  $\mathbb{Z}_2$ fixed point in a
semi-infinite AdS$_5$ bulk. The bulk cosmological constant is
$\Lambda_5=-6/\kf \ell_{\text{AdS}}^2$, where $\ell_{AdS}=M^{-1}_{\text{AdS}}$ is the
AdS$_5$ curvature scale and $\kf=8 \pi G_5$ is the five-dimensional
gravitational coupling constant (in what follows the index AdS
is omitted.)  We note that this was shown to be a
consistent solution of the five-dimensional supergravity theory in
\cite{solution,solution2}.
Although for such a theory to have
a UV completion in quantum gravity (string/M-theory) the presence of additional
dimensions and moduli are certainly of concern, for simplicity we
will assume that such moduli have been fixed by fluxes, or
non-perturbative effects, at a high scale so that they decouple from
the five-dimensional scales that we will consider in what follows.

Using diffeomorphism (gauge)
invariance we are free to choose coordinates to work in the frame where the metric takes the form
\ba
\d s^2=q\mn (x^\mu,y )\d x^\mu \d x^\nu +\d y^2, \label{II ds^2}
\ea
with the brane located at the orbifold fixed point  $y\equiv0$.

The five-dimensional action is
\ba
\label{II 5daction}
S&=&\int \d^5
x \sqrt{-g} \left(\frac{1}{2\kf}\, ^{(5)\!}R-\Lambda\right)
+\int_{\mathcal{M}}\d^4
x\sqrt{-q}\left(\mathscr{L}-\frac{1}{2\kf}\,K\right),
\ea
where
$K_{\mu \nu}$ is the extrinsic curvature of a $y=\text{const}$
hypersurface. In our frame, $K_{\mu \nu}$ is defined by
\ba
\label{K=q'} K_{\mu \nu}(y,x)&=&\frac{1}{2}\partial_y q_{\mu
\nu}(y,x).
\ea
For a brane with positive
tension $\lambda=6/\kf \ell$, and with stress-energy tensor $T\mn$,
the Isra\"el matching condition \cite{Israel:1966rt}
imposes the extrinsic curvature on the
brane to be
\ba
\label{Israel}
K^{\mu\,}_{\, \nu}(y\equiv0)=-\frac{1}{\ell}\, \delta^\mu_\nu - \frac{\kf}{2}\,
 \left( T^{\mu}_{\, \nu}-\frac{1}{3}T \delta^\mu_\nu
\right).
\ea

\subsection{Embedding the inflationary brane}
We now consider the stress-energy tensor for an inflaton
scalar field $\varphi$ confined to the brane:
\ba
\label{Tuv}
T\mn=\partial_\mu \varphi \, \partial_\nu \varphi -\(\frac 12 (\partial
\varphi)^2+V \) q\mn .
\ea
The modified Friedmann equation (Hamiltonian constraint) on the brane is then given by\footnote{
We refer the reader to \cite{BraneFriedEq}, for a careful derivation of the following cosmological equations.}
\be
H^2=\frac{\kf}{3 \ell}\,  \rho \, \(1+\frac{\kf \ell}{12} \rho\),
\ee
where $\rho$ is the energy density on the brane.
Assuming slow-roll inflation and working
to lowest order we have $\bar{T}\mupn\approx -V \delta \mupn$, and the Friedmann equation becomes
\ba
\label{FriedEq}
H^2 = \frac{\kf}{3 \ell}\,  V \, \(1+\frac{\kf \ell}{12} V\).
\ea

We can now focus on the fluctuations to the inflationary brane resulting from the effects of both brane and bulk fields.
The perturbed Einstein equations are
\ba
\label{EinsteinEq}
\delta G\mupn=\frac{\kf \bar z}{\ell}\ \delta T\mupn- \delta E\mupn\, ,
\ea
where $\bar z=\sqrt{1+\ell^2 H^2}=\(1+\frac{\kf
\ell}{6}V\)$, $\delta G\mupn$ is the perturbed Einstein tensor, and
$\delta E\mupn$ the perturbed Weyl tensor.  This result is derived in detail in Appendix
\ref{appendix Mink}.  In particular, there it is shown that the induced value of the Weyl tensor
on the brane is given by \eqref{Efin dS}, \ie
\ba
\label{Efin dS2}
\delta E\mupn=\frac{1}{\ell^2 H^2}\frac{\mathcal{Q}_\nu(\bar z)}{\mathcal{Q}_\nu'(\bar z)}
\frac{\ell\kf}{2}\,\Box \, \delta \Sigma\mupn \, ,
\ea
where $\mathcal{Q}$ is the modified Legendre function, $\nu=\frac12(-1+\sqrt{1+4
(4-\frac{\Box}{H^2})}$ and $\Box$ is the four-dimensional
d'Alembertian on de~Sitter.
As mentioned in the appendix,
the source tensor $\delta \Sigma\mupn$ is given by
\ba
\label{Source2}
\Box \delta \Sigma\mupn=\(\Box \delta T\mupn -\frac 13 \Box
\delta T\, \delta \mupn+\frac 13 \gamma^{\mu \alpha}\delta T_{; \, \alpha \nu}\)
-4H^2\(\delta T\mupn-\frac 14 \delta T \delta\mupn\)\, ,
\ea
and is transverse and traceless.

In the Minkowski limit, $H\rightarrow 0$, one should recover the
result \eqref{MinkowskiWeyl} obtained in \cite{Garriga:1999yh}, which has been
rederived in appendix \ref{appendix Mink}.
The comparison is performed in appendix \ref{appendix check} and
represents a highly non-trivial check of this four-dimensional
effective theory.

\subsection{Anisotropy and an unusual trace coupling}
We note that there are
two main distinctions from standard four-dimensional Einstein gravity arising in the model we have considered\footnote{These modifications are true regardless of whether the slow-roll approximation holds.}.
Firstly, the tensor
$\delta\Sigma\mupn$ presents a source of anisotropic stress which takes the form
$\gamma^{\mu \alpha}D_{\alpha}D_{\nu}\delta T$ and
contributes as an effective shear. We note that this shear is completely
gravitational in origin and will be present even in the absence of
anisotropic matter on the brane.
This effective anisotropy
from the four-dimensional point of view could have
potentially observable consequences, and will be considered in detail in a future publication
\cite{CST}.

In this paper we wish to concentrate on the second
main distinction, namely the peculiar
coupling to the trace. In any four-dimension theory, gravity (\ie gravitons) couples
directly to the stress-energy $T\mupn$, ensuring
gravity to be massless. Although the zero mode graviton is
still massless in this model, the coupling of matter to gravity in (\ref{EinsteinEq}) and (\ref{Efin dS2}) is
modified by an additional term proportional to the
trace.
In the following section we will study the implications of
this unusual trace coupling during inflation, but we first study in
what follows the asymptotic behavior of the effective theory at
short and long wavelengths.

\subsection{UV/IR modifications}

As already mentioned, the effective theory \eqref{EinsteinEq} is only valid at leading order in the
slow-roll parameters. However, it is important to make the distinction between
the time evolution of the background quantities and that of
perturbations. The slow-roll approximation only requires the background
to evolve smoothly and slowly in time (dS approximation), however
perturbations are not constrained by the same requirement and are
free to oscillate very rapidly (such as at short wavelengths) without affecting the validity of the
theory.

At short wavelengths, perturbations evolve much more
rapidly than the background and will therefore live on dS up to a very good approximation.
In other words, small scales cannot feel the slow rolling of the
background
and the approximation that the background is dS will
be a very good one. Fortunately, this will be precisely the regime
we will be interested in, in what follows.
The initial conditions for the curvature perturbation are indeed usually
set by imposing the scalar field to be in the Bunch-Davies vacuum \cite{BD} well inside the horizon.
In the following section we will therefore study how the presence of
the extra-dimension affects the initial conditions.

Since at short wavelengths, the modes are expected to oscillate
rapidly,
we will therefore be interested in the limit where $\Box/H^2 \gg1$.
A good approximation for the operator $r\equiv \frac{\mathcal{Q}_\nu(\bar z)}{\mathcal{Q}_\nu'(\bar
z)}$ appearing in (\ref{Efin dS2}) is then
\ba
\label{r short}
\ \ \ \ r\equiv\frac{\mathcal{Q}_\nu(\bar z)}{\mathcal{Q}_\nu'(\bar
z)}=-\frac{\ell H}{|\nu|} \ \ \ \text{ for } |\nu|  \gg 1 \ \text{ with }\ \ell
H\sim \text{const}\,.
\ea

For completeness, one can also simplify the expression for $r$ in
the long wavelength regime. In this regime, spatial derivatives may
be neglected.
Furthermore, the modes are usually expected to freeze while crossing the
horizon and their time evolution is also negligible. Note
that since the time evolution of these modes is now comparable to
that of the background, such modes will be much more sensitive to
any slow-roll departure from dS. We can thus expect the effective theory \eqref{EinsteinEq}
to be less accurate in that regime, unless the departures from
dS are negligible.
In that regime, the action of the operator $\Box$ on the transverse and traceless
tensor $\Box \delta \Sigma\mupn$ is thus simply $\Box \simeq 8H^2$, \ie $\nu=\frac12 (-1+\sqrt{-15})$.

Using this value of $\nu$, the effective theory can be simplified a step further in both limits
$\ell H\gg 1$ and $\ell H\ll 1$ using the relation
\ba
r\equiv\frac{\mathcal{Q}_\nu(\bar z)}{\mathcal{Q}_\nu'(\bar z)}=
\left\{\begin{array}{lcc}
-\frac  {\ell H}{8} & \text{ for } & \ell H \gg 1\\
\ell^2 H^2 \log \(\frac{\ell H}{2}\) & \text{ for } & \ell H \ll 1
\end{array}\right. \,.
\ea
We also note that $r$ is always negative, even outside these two asymptotic regimes.
As expected the corrections will thus vanish as $\ell H\rightarrow 0$
which can either be seen as the Minkowski limit $H\rightarrow 0$ (with no derivative
terms), or as the small extra-dimension limit ($\ell \rightarrow 0$).
In both cases, the contribution of the Kaluza-Klein tower should indeed vanish.

To summarize, we have reached the important conclusion that if the energy scale of the bulk curvature is comparable or less than the Hubble scale (as can be the case during inflation),
the Kaluza-Klein tower can give a contribution of the same order of
magnitude as the zero mode and should thus be considered with care.


\section{Implications for inflation}
\label{section implications}

In this section we consider density fluctuations on the dS brane
to which the inflaton scalar field is confined.

We will work in longitudinal gauge where the metric for scalar metric perturbations is
\ba
\label{long gauge}
\d s^2&=&a(\tau)^2 \left[ -(1-2\Phi)\d \tau^2+ (1+2\Psi) \eta\mn \d x^\mu \d
x^\nu \right] \\
\varphi&=&\varphi_0(\tau)+\delta \varphi(\tau,x^\mu)\,,
\ea
where we will work using conformal time $d\tau^2=a^{-2} dt^2$.
In the absence of anisotropic stress, the Einstein equations imply $\Phi=\Psi$, so that there is
one scalar degree of freedom associated to the metric. As mentioned previously,
we do expect anisotropy contributions to be present in this
scenario, however they are studied in detail in \cite{CST} and we
focus in what follows in the other types of corrections arising from
the Weyl tensor.

Following the usual four-dimensional prescription \cite{Inflation},
it is useful to introduce the gauge invariant
curvature perturbations on uniform density hypersurfaces as\footnote{Note that at long wavelength, the above definition of the curvature
perturbation coincides with that on comoving hypersurfaces $\zeta_\varphi=\Psi -a H \frac{\delta \varphi}{\dot
\varphi_0}$, and that on uniform effective energy density hypersurfaces
 $\zeta_{eff}=\Psi -a H \frac{\delta \rho_{\text{eff}}}{\dot
\rho_{\text{eff}}}$, where $\rho_{\text{eff}}$ is the total
effective energy density as seen by the metric. Since we are
interested in the power spectrum and spectral index at late time, we are free to use
any of these three definitions.}
\ba
\zeta=\Psi -a H \frac{\delta \rho}{\dot \rho}
\ea
where a dot represents the derivative with respect to the conformal
time $\tau$.

In what follows, we will analyze how the evolution of the curvature perturbation $\zeta$
differs from the standard four-dimensional case. In
particular, we will show that the dynamics of $\zeta$ and $\delta
\varphi$ are both governed by the same set of coupled differential
equations as in the standard four-dimensional case.
We will thus argue that the only
possible departure from the usual result can only arise from a
modification of the initial conditions.
We will then study in detail the initial
conditions to be imposed at short wavelengths on the curvature perturbations.
Although the gravitational potential $\Psi$ is
greatly enhanced by the higher curvature effects at
short wavelengths, the gravitational backreaction still remains
negligible in this regime, unaffecting  the initial condition
of $\zeta$ and thus its time evolution.

\subsection{Dynamics}

In what follows we study the dynamics of the perturbations. Given the isotropy of the model, there
are only two independent variables, which we will take to be $\delta \varphi$ and
$\zeta$. These two variables are coupled via the Einstein equations.
We first observe that because the scalar field is confined to the
brane, its equation of motion will remain unaffected by the presence of the
extra-dimension.
Furthermore, the Weyl tensor defined in \eqref{Efin dS2} is
traceless. The trace of the Einstein equation \eqref{EinsteinEq} will therefore be similarly
unaffected by the extra-dimension, outside of ``background
effects''\footnote{By ``background effects" we refer to the effect
that comes uniquely from the unusual Friedmann equation \eqref{FriedEq} and are
independent of the Weyl tensor.} which have been studied extensively in the literature
\cite{braneinflation}.

Both the scalar field and the trace equations:
\ba
\delta T^\mu_{\, \nu \, ;\,  \mu}=0\quad \text{ and }\quad \delta R=-\frac{\kf \bar z}{\ell}\, \delta T
\ea
 will thus provide
two dynamical equations for $\delta \varphi$ and $\zeta$ which are identical to
their usual four-dimensional counterpart in standard inflation. These two dynamical equations govern
the entire dynamics of the system and will provide
two coupled equations for $\zeta$ and $\delta \varphi$.
In other words, the fact that the bulk is empty
(implying that the Weyl tensor is traceless) and that the scalar field is confined to the brane ensure
that the equations of motion for the curvature perturbations and the scalar field will be unaffected
by the extra dimension outside of ``background effects".

Thus, the only way the extra-dimension can interfere in the evolution of the perturbations is through the setting of the initial
conditions.
We have here a set of two coupled second order equations for $\zeta$ and
$\delta \varphi$, or equivalently a fourth order equation for
$\zeta$. We will thus require four initial conditions to fix $\zeta$
entirely.

Two of these initial conditions can be fixed by the usual
requirement that the scalar field is in the
Bunch-Davies vacuum \cite{BD} well inside the horizon. But to fix the
two remaining ones, we should make use of the constraint equations. We
emphasize that the constraint equations (the Hamiltonian constraint $G^0_{\, 0}$ and the momentum density constraint  $G^0_{\, i}$)
are affected by the Weyl tensor and will thus give a different
relation than in the standard case. In principle, we
expect the initial conditions for the
curvature perturbation to be affected at short wavelengths, which might in turn
affect its evolution and the observed power spectrum. This
possibility is explored in what follows.

\subsection{Constraints and Initial Conditions}

\subsubsection{Standard Chaotic Inflation}

We begin by analyzing the constraints and initial conditions for the case of standard chaotic inflation.
As mentioned previously, two of the initial conditions are fixed by imposing the scalar field $\delta \varphi$ to be in the Bunch-Davies
vacuum at short wavelength, \ie for $k \tau \gg 1$,
\ba
\label{BDvacuum}
\delta \varphi _k\sim -\frac{\ell}{a \kf} \frac{e^{-i k
\tau}}{\sqrt{2k}}\(1-\frac{i}{k \tau}\)\, .
\ea
In terms of the energy density perturbation, this implies, to
leading order in the slow-roll parameters,
\ba
\frac{\delta \rho_k}{\dot \rho}=\frac{\delta \varphi_k}{\dot
\varphi_0}-\frac{\dot{\delta \varphi_k}}{V'}
\approx-\frac{\dot{\delta \varphi_k}}{V'}\,,
\ea
where the second equality holds at short wavelengths.

In order to fix the two remaining initial conditions, we use the
constraints equations $G^0_{\, 0}$ and $G^0_{\, i}$ to express the metric perturbation
$\Psi$ in terms of $\delta \varphi$. This gives the
following relation in the slow-roll regime
\ba
\Psi_k=\kq \frac{\dot \varphi_0}{ 2 k^2} \, \delta \dot \varphi_k\,, \nn
\ea
where $\kq=\kappa_5^2/ \ell$ is the four-dimensional gravitational coupling constant.
At short wavelengths, the contribution from the gravitational potential $\Psi$ is thus
completely negligible compared to the contribution of the scalar
field  when setting the initial conditions for
$\zeta$: $\Psi_k \propto \frac{\delta \dot \varphi_k}{ k^2}  \ll \frac{a H}{V'} \dot{\delta \varphi_k}$, so
that  $\zeta_k\approx \frac{a H}{V'} \, {\delta \dot \varphi_k}$.
This is
a simple consequence of the nature of inflation for which the
gravitational backreaction is very weak. On small scales,
the perturbations are unable to distinguish between an expansion
driven by a scalar field or by a cosmological constant. Since a
cosmological constant would have very little effects on
scalar perturbations (beside ``background effects"), it is no surprise that
the gravitational potential $\Psi$ plays little role in setting the
initial conditions at small scales.

We now consider how this argument is modified in the context of
brane inflation.

\subsubsection{Inflation on the Brane}

As mentioned previously, the constraints equations  $G^0_{\,0}$ and $G^0_{\, i}$
are affected by the presence of the non-local corrections embedded in the
Weyl tensor. They will therefore give rise to a slightly modified
relation between $\Psi$ and $\delta \varphi$. Using the Einstein
equation \eqref{EinsteinEq} in the gauge \eqref{long gauge}, we indeed
have the modified relation:
\ba
\Psi_k  =  \frac{\kf}{\ell} \frac{\dot \varphi_0}{12 a^2 H^2 k^2}
\((5rk^2 +6 \bar z a^2 H^2 )\, \delta \dot \varphi_k
-6 r a H \delta \ddot \varphi_k+3 r \delta \dddot \varphi_k
\)\,,
\ea
which holds at short wavelengths and in the slow-roll approximation.
The derivation of this result is detailed in appendix \ref{appendix
short}.
For a scalar field in the Bunch-Davies vacuum \eqref{BDvacuum}, the
behaviour of the gravitational potential is thus
modified to
\ba
\label{Relation 2}
\Psi_k \approx r\,  \frac{\kf}{\ell} \frac{\dot \varphi_0}{6 a^2 H^2} \delta \dot
\varphi_k\,.
\ea
The gravitational potential is thus enhanced by the
non-local curvature corrections encoded in $\delta E\mupn$. {\em A priori}
this could imply significant changes in the initial conditions for $\zeta$,
which will then propagate to late times and freeze out.
However, at short wavelengths, $|\nu|^2 \approx 4 k \tau$, so that using the relation \eqref{r short} in the limit $k \tau
\gg1$, $r\approx -\ell H /2\sqrt{k \tau}$ (Cf. appendix \ref{appendix short}).

Due to this overall factor, the contribution from $\Psi$ is thus still
negligible compared to that of the scalar field in the definition
of $\zeta$: $\Psi_k \propto  \delta \dot
\varphi_k / \sqrt{k \tau} \ll \delta \dot \varphi_k $, so that the
initial conditions for $\zeta$ will still be dominated by $\zeta_k\approx \frac{a H}{V'} \, {\delta \dot
\varphi_k}$, as was the case in the standard chaotic inflation
scenario.

We can therefore conclude that despite the fact that the
gravitational potential $\Psi$ is modified at short
wavelengths in this scenario, its overall contribution to the initial
conditions for $\zeta$ remains negligible. The initial conditions for
$\zeta$ and $\delta \varphi$ will thus remain unaffected by the
presence of the extra-dimension, and since their evolution is
governed by a set of equations which is identical to that in four
dimensions, we must conclude that no signature of the
extra-dimension will be present in the power spectrum and the
spectral index to leading order in the slow-roll parameters, except for
the possibility of  ``background effects".

\subsubsection*{Remarks on Anisotropy}

In order to focus on the key point of the argument, anisotropy has
been neglected in the previous development.
It is however
important to notice that the previous result is independent of this
assumption and is completely generalizable to the case where $\Phi \ne
\Psi$.
At long wavelengths, the anisotropic part of the Weyl tensor
vanishes, so that one should recover $\Phi =
\Psi$.
At short wavelengths, however,
the relation between the two gravitational potentials can be
derived from the last constraint $G^i_{\, j}$ (with $i\ne j$) to
give
\ba
\Phi=\Psi - r \frac{\kf}{\ell}\frac{\dot \varphi_0}{3 a^2 H^2}\,
\delta \dot \varphi\,.
\ea
Despite this difference, the relation between $\Psi$ and $\delta
\varphi$ remains unchanged and is still given by \eqref{Relation 2}, (Cf. appendix \ref{appendix short}).
The contribution from the gravitational potential in the initial
conditions for $\zeta$ will therefore remain negligible.

Furthermore, as seen in the appendix  \ref{appendix short}, at short
wavelengths the equation for the scalar field is similarly unaffected
by the anisotropy \eqref{ConserEqAnisotropy}. Since the anisotropy
disappears at long wavelengths, and the transition between the
short-wavelength oscillating modes and the long wavelength frozen
modes occurs almost instantaneously, the equation of motion for the
scalar field will be unaffected by the extra-dimension (through the Weyl
tensor) during its entire evolution. The argument given previously
will therefore remain valid even in the presence of anisotropy and
no signature to the power spectrum will be present to leading order
in the slow-roll parameters, besides that of the background effects.

\section{Conclusion}
\label{section conclusion}

In this paper, we have developed a formalism capable of tracking
non-local effects of the KK continuum at the perturbed level on a
single de~Sitter brane while preserving general covariance on the brane. As a consistency check, this formalism was shown to reproduce the
correct Minkowski limit developed in \cite{Garriga:1999yh}.
We argue that to leading order in the slow-roll parameters, this
effective theory provides a good framework to study models of
braneworld inflation for which the inflaton scalar field is confined
to the brane.

In particular, we use this analysis to study the power spectrum of
curvature perturbations.
Although non-local terms typically present in braneworld scenarios
can be very important at short scales, a close analysis to their
effects during a standard model of inflation
shows that the corrections to the power spectrum are completely
negligible in the slow-approximation besides any possible background effect.

This is in complete agreement with the results of \cite{Koyama:2007as}, which
appeared during the latest stages of this work. Both
studies use similar approximations (slow-roll) but
different methods to derive very analogous results.
We may point out that the present analysis does not rely on any
numerical computations and is thus very easily reproducible. Another
advantage of our analysis resides in its
covariant derivation. It can thus very easily be extended to the study of the
tensor modes or to non-inflationary scenarios which left for a
further study.

\acknowledgments
We would like to thank Sera Cremonini, Nemanja Kaloper, and Andrew
Tolley for useful discussions.
CdR and SW acknowledge support from the Natural
Sciences and Engineering Research Council of Canada.
SW would also like to thank UC-Davis and the University of
Michigan-MCTP for hospitality and financial support during the
completion of this work.
Research at Perimeter Institute for Theoretical Physics is supported
in part by the Government of Canada through NSERC and by the
Province of Ontario through MRI.



\appendix

\section{Effective theory on a Minkowski and de~Sitter brane}
\label{appendix Mink}
\subsection{Four-dimensional decomposition}

We consider the five-dimensional theory \eqref{II 5daction}, and
work in the gauge \eqref{II ds^2} where the brane is fixed at $y=0$.

The electric part of the five-dimensional Weyl tensor
$E_{\mu \nu}$ is defined as:
\ba
\label{II E=C}
E_{\mu \nu} (y,x)\equiv\,
^{(5)\!}C^{y}_{\phantom{y}ayb}\, q^a_{\, \mu} q^b_{\, \nu}.
\ea
From
the properties of the five-dimensional Weyl tensor $\,
^{(5)\!}C^{a}_{\phantom{a}bcd}$, we can easily check that $E_{\mu
\nu}$ is traceless with respect to $q_{\mu \nu}$. In the frame
(\ref{II ds^2}), using the five-dimensional Einstein equation
$G_{ab}=-\kf \Lambda g_{ab}$,
 the expression for $E\mn$ is
\ba
E^\mu_{\,\nu}=\phantom{.} ^{(5)\!}R^y_{\phantom{y}\alpha y  \nu} q^{\alpha \mu}
+\frac{1}{\ell^2}\, \delta^\mu_\nu
=-\partial_y K^\mu_{\,\nu}-K^\mu_{\,\alpha}K^\alpha_{\,
\nu}+\frac{1}{\ell^2}\delta^\mu_{\, \nu}\label{II K'} \,.
\ea
It has been shown in the literature, that from a four-dimensional
brane perspective, the projected Weyl tensor on the brane is the only quantity mediating
between the brane and the bulk.
Indeed, the projected Ricci tensor on
the brane can be expressed as
\begin{eqnarray}
R_{\mu \nu } (x)= \frac{\kf}{\ell} \left( T_{\mu \nu }-\frac{1}{2}T q_{\mu \nu }\right)
-\frac{\kappa_5^4}{4}\left( T_{\mu }^{\alpha }T_{\alpha \nu
}-\frac{1}{3}TT_{\mu \nu }\right) -E_{\mu \nu }(0,x).  \label{II
Ricci1}
\end{eqnarray}
The evolution of the Weyl tensor through the bulk is governed by the
equation:
\ba
\partial_y E^\mu_{\, \nu}
&=&
2K^{\alpha}_{\nu} E^{\mu}_{\alpha}-\frac{3}{2} K
E^{\mu}_{\nu}-\frac 1 2 K^{\alpha}_{\beta} E^{\beta}_{\alpha}
\delta^{\mu}_{\nu} \label{II E'}
+C^{\mu}_{\phantom{\mu} \alpha
\nu
\beta}K^{\alpha \beta}
+(K^{3})^\mu_{\, \nu}\\
&& -\frac{1}{2}D^{\alpha} \left[
D^{\mu} K_{\alpha \nu}+D_{\nu} K^{\mu}_{\alpha}-2 D_{\alpha}
K^{\mu}_{\nu}
\right]\,,\nn
\ea
where $(K^{3})^\mu_{\, \nu}$ are some cubic terms in the
traceless part of the extrinsic curvature whose exact form will not be relevant for the
purpose of this study (it is enough to note that this term vanishes
if $K\mupn \sim \delta \mupn$).
The previous results have first been developed in
ref.~\cite{Shiromizu:1999wj}.

\subsection{Linearized gravity on a Minkowski brane}
We now use this formalism to recover the well-know result of Garriga
and Tanaka in ref.~\cite{Garriga:1999yh} that first studied
linearized gravity around a  Minkowski brane, expressing explicitly
the KK tower and their effect on the brane.

For an empty brane, the background bulk geometry is
exactly AdS: $\bar q\mn(y,x)=e^{-2
y/\ell}\bar{\gamma}\mn(x)$, where $\bar\gamma\mn$ is the
metric on the brane, here $\bar \gamma\mn=\eta\mn$. Since
the background in conformally invariant, the Weyl tensor vanishes
at that order. From the evolution equation \eqref{II K'}, together with the
boundary condition \eqref{Israel}, the extrinsic curvature is thus simply
$\bar{K}\mupn(y,x)=-\frac{1}{\ell}\delta\mupn$.

We now consider small fluctuations induced by matter on the brane
$T\mupn=0+\delta T\mupn$, so that the extrinsic
curvature and the Weyl tensor become:
\ba
E\mupn&=&0+\delta E\mupn \nn \\
K\mupn&=&\bar K \mupn+\delta K\mupn \nn \\
\delta K\mupn(y\equiv0)&=&-\frac{\kf}{2}(\delta T\mupn-\frac 13 \delta T \delta
\mupn)\nn .
\ea
At the linearized level, the evolution equations (\ref{II K'},\ref{II
E'}) for the extrinsic curvature and the Weyl tensor thus simplify
to
\ba
\label{dK 1}
\partial_y \delta K\mupn & =&  - \delta E\mupn +\frac 2 \ell \delta K\mupn
\\
\label{dE 1}
\delta_y \delta E\mupn &=&\frac 4 \ell \delta E\mupn
+e^{2y/\ell}\(\Box \, \delta K\mupn-\bar\gamma^{\mu\alpha}\, \delta K_{;\, \alpha
\nu}\),
\ea
where $\Box $ is the four-dimensional d'Alembertien in
Minkowski: $\Box=\gamma^{\alpha \beta}D_{\alpha}D_{\beta}$.
Combining these two equations we get the second order differential
evolution equation for the Weyl tensor:
\ba
\partial_y^2 \delta E\mupn=\frac 8 \ell \partial_y \delta E
\mupn-\frac{16}{\ell^2} \delta E\mupn - e^{2y/l} \Box \, \delta
E\mupn,
\ea
which has the simple solution:
\ba
\delta E\mupn (y,x)= e^{4y/\ell}\(I_0 (e^{y/\ell}\ell \sqrt{-\Box}) \, A\mupn(x)+K_0 (e^{y/\ell}\ell \sqrt{-\Box})\,
B\mupn(x)\),
\ea
where $K_0, I_0$ are the two independent Bessel functions, and $A\mupn,\,
B\mupn$ are two ``integration constants" to be determined. One of
them is imposed by the requirement that the five-dimensional metric
(and thus the Weyl tensor) remains finite at spatial infinity ($y\rightarrow
\infty$). This requirement sets $A\mupn$ to zero. The second tensor
$B\mupn$ is fixed by the Isra\"el matching condition on the brane
\eqref{Israel}. Using the relation \eqref{dE 1}, one has:
\ba
\Box \, \delta K\mupn(0)-\bar\gamma^{\mu\alpha}\, \delta K_{;\, \alpha
\nu}(0)=-\frac{\kf \Box}{2}\delta \Sigma\mupn &\equiv&
-\frac{\kf \Box}{2}\(\delta T\mupn -\frac{1}{3}\delta T
+\frac{1}{3 \Box} \bar \gamma ^{\mu \alpha}\delta T_{; \alpha
\nu}\)\nn \\
&=&-\sqrt{-\Box} \, K_1(\ell \sqrt{- \Box})\, B\mupn \,,
\ea
so that the Weyl tensor on the brane is
\ba
\label{MinkowskiWeyl}
\delta E\mupn (0,x)=-\frac{\kf}{2}\, \sqrt{- \Box}\,
\frac{K_0 (\ell \sqrt{-\Box})}{K_1(\ell \sqrt{- \Box})} \,
\delta \Sigma\mupn \,.
\ea
The modified Einstein equation on the brane is therefore:
\ba
\delta R\mupn = \frac{\kf}{\ell}\(\delta T\mupn -\frac 1 2 \delta T \delta
\mupn\)+\frac{\kf}{2}\, \sqrt{- \Box}\,
\frac{K_0 (\ell \sqrt{- \Box})}{K_1(\ell \sqrt{- \Box})} \,
\delta \Sigma\mupn,
\ea
which is precisely the result obtained by Garriga and Tanaka in
ref.~\cite{Garriga:1999yh}. Motivated by this result, we can now use
the same formalism to compute the KK tower for a dS brane, which
will be useful if we are to study inflation on the brane.

\subsection{Linearized gravity on a de~Sitter brane}
\label{appendix dS}

Still working in the frame \eqref{II ds^2}, where the brane location
is fixed at $y=0$, we now consider a positive cosmological constant
on the brane (or equivalently, assume that the brane tension is
larger than its conical value $\lambda$), so that for the background
\ba
T\mupn=-V \delta \mupn .
\ea
The modified Friedmann equation is then given by
\ba
H^2=\frac{\kf}{3 \ell}\,  V \, \(1+\frac{\kf \ell}{12} V\),
\ea
and at
the perturbed level, the Einstein equation on the brane will be:
\ba
\delta R\mupn=\frac{\kf}{\ell}\(1+\frac{\kf \ell}{6}V\)\(\delta T\mupn-\frac 12 T
\delta\delta\mupn\)- \delta E\mupn.
\ea

Solving the evolution equation \eqref{II K'} with the boundary
condition imposed by the Isra\"el matching condition \eqref{Israel},
we thus have:
\ba
\bar K \mupn(y) = -\frac{1}{\ell}k(y)\delta\mupn=-\frac 1 \ell \coth (\bv-\frac y \ell) \delta \mupn,
\ea
where $\bv$ is an integration constant fixed such that the right boundary condition is recovered on the
brane \ie $\bar K \mupn(0)=-\frac 1 \ell (1+\frac{\kf \ell}{6}V)$, imposing $\bv=\text{arccoth} (1+\frac{\kf \ell}{6}V)$.
The metric profile is therefore:
\ba
\partial_y \bar q\mn=2\bar K\mn\ \ \  \Rightarrow \ \ \ \bar q\mn (y,x)= \frac{ \sinh^2 (\bv-y/\ell)}{\sinh^2 \bv}\ \bar{\gamma}\mn(x),
\ea
where here again $\bar{\gamma}\mn(x)$ is the induced metric on the
brane. We choose here to work in conformal time, so that $\bar \gamma\mn = a(\tau)^2
\eta\mn$, with $H=\dot a /a^2$.
 In the limit $V\rightarrow
0$, $\bv \rightarrow +\infty$, we recover the result for a Minkowski
brane: $k(y)= 1$ and $\bar q\mn (y,x)=e^{-2y/\ell}\,
\bar{\gamma}\mn(x)$.

We may notice at this point, that in this gauge,
spatial infinity is reached
at finite value of $y$, \ie at $y=\bar y_\infty =\ell \bv >0$. The
perturbations perceived on the brane will depend on the boundary
conditions imposed at spatial infinity. The mode functions should be normalized over the region $0<y<\bar y_\infty $, which
will require them to fall as $y\rightarrow \bar y _\infty$ or behave as plane
waves. Note that at this point, the extrinsic curvature is infinite
$k(y)\rightarrow \infty$ as $y\rightarrow \bar y_\infty $.

\subsubsection*{Weyl tensor}

As seen earlier, the key point in obtaining the KK tower at the
perturbed level, is to find the expression for the Weyl tensor.
We proceed as for the Minkowski brane, and the evolution equations
(\ref{II K'}, \ref{II E'}) for the extrinsic curvature and the Weyl tensor at the perturbed
level are:
\ba
\partial_y \delta K\mupn&=&- \delta E\mupn+\frac 2\ell k(y) \delta
K\mupn \\
\label{dE 2}
\partial_y \delta E \mupn&=&\frac 4 \ell k(y) \delta E\mupn+\frac{1}{\ell ^2 H^2 \sinh^2 (x-\frac y l)}
\left[\hat O \delta K\mupn-\hat O ^\mu_\nu\delta
K\right],
\ea
with the operator $\hat O^\mu_\nu=\bar{\gamma}^{\mu \alpha} D_{\alpha} D_{\nu}-H^2
\delta\mupn$, and $\hat O=\hat O^\alpha_\alpha=\Box-4H^2$.
Combining these two equations, and using the variable
$z=k(y)>1$, we get the evolution equation for the Weyl tensor
\ba
\partial_y^2\delta E\mupn&=&\frac 8 \ell k(y)\partial_y\delta E\mupn
+\frac 4 \ell \(k'(y)-\frac 4\ell k^2(y)\)\delta E\mupn -
\frac{1}{\ell^2 H^2}\hat O \delta E\mupn\nn \\
(z^2-1)^2\partial_z^2 \delta E\mupn&=&6z (z^2-1)\partial_z \delta
E\mupn-4 (3z^2+1)\delta E\mupn-(z^2-1)\frac{1}{H^2}\hat O \delta
E\mupn \, .
\ea
The general profile of the Weyl tensor through the bulk can be expressed of the
form
\ba
\delta E\mupn(z)=(z^2-1)^2 \left[P_\nu(z) C\mupn+Q_\nu(z)
D\mupn\right],
\ea
where $P_\nu$ and $Q_\nu$ are the Legendre functions, and $\nu=\frac 12 \(-1+\sqrt{1-\frac 4{H^2}\hat
O}\)$. The profile of the Weyl tensor is thus drastically modified
compared to the situation in Minkowski.
The integration tensors $C\mupn$ and $D\mupn$ are fixed by the
requirement that the Weyl tensor falls at spatial infinity
$z\rightarrow \infty$ or behaves as plane waves,
and by the Isra\"el matching condition.

The constraint at spatial infinity, together with the
requirement that the Weyl tensor is real, sets its general profile
to be
\ba
\delta E\mupn(z)=  (z^2-1)^2 \left[ \mathcal{Q}_\nu
(z)+\mathcal{Q}_{\nu^\star}(z)
\right]\, C\mupn \, ,
\ea
where $\mathcal{Q}_\nu$ is the modified Legendre function:
\ba
\mathcal{Q}_\nu(z)=z^{-\nu-1} {}_2F_1\(\frac{1+\nu}{2},\frac{2+ \nu} 2,\frac {3+2\nu}{2},
z^{-2}\)
+\text{cc}\,,
\nn
\ea
and $_2F_1$ is the Hypergeometric function of second kind.

Finally, the second integration constant $C\mupn $ is fixed by the
Isra\"el matching condition on the brane \eqref{dE 2} which reduces to
\ba
\label{Israel 2}
 \mathcal{Q}_\nu'(\bar z)\, C\mupn
=\frac{1}{\ell^6 H^6}\frac{\ell\kf}{2}\Box \delta \Sigma\mupn\, ,
\ea
where $\Box$ is the now d'Alembertian in de~Sitter space, $\Box=\bar{\gamma}^{\mu
\nu}D_\mu D_\nu$,
and $\bar z$ is the value of $z$ on the brane: $\bar z=\(1+\frac{\kf \ell
}{6}V\)=\sqrt{1+\ell^2H^2}$.
The source tensor $\delta \Sigma\mupn$ is given by
\ba
\label{Source}
\Box \delta \Sigma\mupn&=&\frac{2}{\kf} \left[\hat O \delta K\mupn(0)-\hat O^\mu_\nu\, \delta
K(0)\right]\nn\\
&=&\(\Box \delta T\mupn -\frac 13 \Box
\delta T\, \delta \mupn+\frac 13 \gamma^{\mu \alpha}\delta T_{; \, \alpha \nu}\)
-4H^2\(\delta T\mupn-\frac 14 \delta T \delta\mupn\).
\ea
One can check that $\delta \Sigma\mupn$ is indeed transverse and traceless
with respect to the background de~Sitter metric $\bar \gamma\mn$.
For any arbitrary function $f(\Box)$, the tensor $f(\Box)
\delta\Sigma \mupn$, will thus also remain transverse and traceless.

Using this boundary condition, the Weyl tensor on the brane is thus of the form:
\ba
\label{Efin dS}
\delta E\mupn (\bar z)=\frac{1}{\ell^2 H^2}\frac{\mathcal{Q}_\nu(\bar z)}{\mathcal{Q}_\nu'(\bar z)}
\frac{\ell\kf}{2}\,\Box \, \delta \Sigma\mupn \, ,
\ea
with the perturbed Einstein equation
\ba
\delta G\mupn = \frac{\kf}{\ell}\, \bar z\, \delta T\mupn-\delta
E\mupn\,.
\ea

\section{Minkowski versus de~Sitter brane}
\label{appendix check}

The aim of this appendix is to show that the dS effective
theory reduces to that of Minkowski in the limit $H \rightarrow 0$.
The main difficulty in taking this limit resides in the fact that
$\nu$ is then ill-defined: $\nu\sim \sqrt{-\frac{\Box}{H^2}}\,$.
We thus first consider the limit $\bar z=\sqrt{1+\ell^2 H^2}\rightarrow
1$, with fixed $\nu$.
\ba
\frac{1}{\ell^2H^2}\frac{\mathcal{Q}_\nu(\bar z)}{\mathcal{Q}_\nu'(\bar
z)}&=&\left[\frac 1 \nu+\(\bar \Gamma +\bar \Gamma(\nu)+\log\(\frac{\ell
H}{2}\)\)\right]\nn\\
&-&\frac{\ell^2 H^2}{8\nu}\Bigg[
2 \Big(\nu  \left(2\nu (1+\nu )\bar \Gamma ^2  +\nu(\nu -1) (1+\log 4)  -2   (1+\nu) (\log 4 \nu +\nu -2)\bar \Gamma-4 \log 2+1\right)\nn \\
&&+2 \nu
   (1+\nu) \log (H\ell ) (\nu \log (H\ell ) + \nu(-1+2 \bar \Gamma -\log 4) +2)+2\Big)\nn \\
  && +\nu  (1+\nu) \left(\nu  \left(\log 4-2 \bar \Gamma(\nu)
   \right)^2+4 (2  \nu \log (H\ell ) +\nu (2 \bar \Gamma -1)  +2)
   \bar\Gamma(\nu)\right)\Bigg]\nn\\
   & +&\cdots \nn \,,
\ea
with $\bar \Gamma$ the Euler number ($\bar \Gamma\sim 0.57$), and $\bar \Gamma(\nu
)$ the digamma function.

Using this expansion, we may now set $\nu= \sqrt{-\frac{\Box}{H^2}}$
and take the limit $H\rightarrow 0$ to obtain:
\ba
\label{Min limit}
\frac{1}{\ell^2H^2}\frac{\mathcal{Q}_\nu(\bar z)}{\mathcal{Q}_\nu'(\bar
z)}=\left[\bar \Gamma+\log\(\frac{\ell k}{2}\)\right]
-\frac{\ell^2k^2}{4}\left[1+2\(-1+\bar \Gamma+\log\(\frac{\ell
 k}{2}\)\)\(\bar \Gamma+\log\(\frac{\ell k}{2}\)\)\right]+\cdots
 \,,\quad\quad
\ea
where for simplicity, we used the notation $k=\sqrt{-\Box}$.

We can see that this result corresponds to a derivative expansion in
the limit where $H=0$. We can therefore check that this expansion is
the same as the one obtained in the Minkowski case for which (cf. eq~\eqref{MinkowskiWeyl})
\ba
\delta E_{\text{Min}}\,\mupn=-\frac{\kf}{2}\, k\,
\frac{K_0 (\ell k )}{K_1(\ell k)} \,
\delta \Sigma\mupn \,,
\ea
where both source terms $\delta \Sigma\mupn$ clearly coincide when
$H=0$. One can easily check that expanding $-\frac{1}{\ell k} \frac{K_0 (\ell k )}{K_1(\ell
k)}$ to second order in $k$ reduces precisely to the previous result
\eqref{Min limit}, confirming that we recover the Minkowski limit as
$H\rightarrow 0$, at least to leading orders in $k$. The comparison
can be continued order by order in $k$ to check the validity of the
theory to all order in derivatives in the Minkowski limit.


\section{Short-Wavelength regime}
\label{appendix short}
We compute in this section the source term \eqref{Source2} using the
perturbed contribution from the stress-energy tensor \eqref{Tuv}.
At short wavelengths and to leading order in the slow-roll parameters,
its expression is simply of the form
\ba
\label{SourceShort}
S\mupn\equiv\Box \delta \Sigma\mupn=\frac{\dot \varphi_0}{3 a^2}
\(\begin{array}{cc}
-5 \nabla ^2 \delta \dot \varphi+3 \delta \dddot \varphi
& \delta \ddot \varphi^{, i} -3 \nabla^2 \delta \varphi^{, i}\\
-\delta \ddot \varphi_{, j} +3 \nabla^2 \delta \varphi_{, j} &\ \
\(\nabla ^2 \delta \dot \varphi- \delta \dddot
\varphi\)\delta^i_{\,j}+ 2 \delta \dot \varphi ^{\, ,i}_{\,\, ,j}
\end{array}\)
\,,
\ea
with $\nabla^2$ the three-dimensional Laplacian
$\nabla^2=\partial^i\partial_i$.

The operator of $\Box$ on this tensor is highly non-trivial, however
in the short-wavelength regime, the time-dependance of $\delta
\varphi$ is given by \eqref{BDvacuum}, so expanding $\delta \varphi$
in Fourier modes, on has for each mode,
$\delta \varphi_k\sim \frac{e^{-i k \tau}}{a}$. Using this behaviour,
we get at short wavelength,
\ba
\Box S\mupn\simeq 4\frac{H}{a}\partial_\tau S\mupn \simeq -4 i
\frac{k H}{a} S\mupn\,.
\ea
This relation can be used to evaluate the action of the operator $r\equiv \frac{\mathcal{Q}_\nu(\bar z)}{\mathcal{Q}_\nu'(\bar
z)}$ on this source term. Recalling $\nu=\frac12(-1+\sqrt{1+4
(4-\frac{\Box}{H^2})}$. At short wavelengths, the action of $\nu$ on
the source term $S\mupn$ will therefore be
$\nu^2 \simeq -\frac{\Box}{H^2}\simeq  4 i
\frac{k}{aH}\simeq -4 i k\tau$ as $k\tau \gg1$, where we used
$a=-1/H\tau$ in the slow-roll regime.

Using the asymptotic behaviour \eqref{r short}, the action of $r$ on
the source term $S\mupn$
will therefore be
\ba
r\rightarrow -\frac{\ell H }{|\nu|}\simeq -\frac{\ell H}{2 \sqrt{k
\tau}} \quad \text{ for }\quad k \tau \gg1\,.
\ea

We may now study the modified Einstein equation
\ba
\delta G\mupn=\frac{\kf \bar z}{\ell }\, \delta T\mupn -\frac{\ell \kf }{2\ell^2 H^2
} r \, S\mupn \,,
\ea
at
the perturbed level. The relation between the two gravitational
potentials $\Psi$ and $\Phi$ can be derived from the $(ij)$ (with $i\ne
j$) component of this equation which reads
\ba
\label{Psi Phi}
\partial^i\partial_j \left[
3 a^2 H^2 \(\Phi-\Psi\)+r \, \dot \varphi_0 \frac{\kf}{\ell} \delta
\dot \varphi\right]=0\,.
\ea
This expression can be used in the two other constraints. In
particular the momentum constraints is of the form
\ba
\label{mom constraint}
\partial_i \left[12 a^3H^2 \Phi +12 a^2 H^2 \dot \Psi +6 \frac{\kf \bar z}{\ell}\dot \varphi_0
a^2 H^2 \delta \varphi + r \, \frac{\kf}{\ell} \dot \varphi_0\(3 k^2
\delta \varphi+\delta \ddot\varphi\)\right]=0\,,
\ea
and the Hamiltonian constraint is
\ba
\label{Ham constraint}
-12 a^2 H^2 \(k^2 \Psi+3 a^2 H^2 \Phi +3 a H \dot \Psi\)
&+&6\frac{\kf \bar z}{\ell}\dot \varphi_0 a^2 H^2 \(\delta \dot
\varphi-3 a H \delta \varphi\) \\
&+&r\, \frac{\kf}{\ell}\dot \varphi_0 \(5 k^2 \delta \dot \varphi-9 a
H
k^2 \delta \varphi-9 a H \delta \ddot \varphi +3 \delta \dddot
\varphi\)=0\,.\nn
\ea
Using the anisotropy relation \eqref{Psi Phi} and the two previous
constraints, we may express $\Psi$ in terms of $\delta \varphi$ as
\ba
\Psi=\frac{\kf}{12\ell}\frac{\dot \varphi_0}{a^2H^2}\(
6 \bar z a^2 H^2 \delta \dot \varphi+r \(6 k^2 \delta \dot \varphi-6
a H \delta \ddot \varphi+3 \delta \dddot \varphi\)\)\,.
\ea
This expression for $\Psi$ has been derived while taking into
account the anisotropy \eqref{Psi Phi}, but we may point out that
the same expression exactly would have hold have we made the
hypothesis $\Phi=\Psi$ in \eqref{mom constraint} and \eqref{Ham
constraint}.

In the absence of the Weyl term ($r=0$), we recover the usual
four-dimensional result. At short wavelengths, the expression for
$\Psi$ is dominated by
\ba
\Psi\simeq r\, \frac{\kf}{4\ell}\frac{\dot \varphi_0}{a^2H^2}\(
2 k^2 \delta \dot \varphi+ \delta \dddot \varphi\)\,.
\ea

Furtheremore, the conservation of energy for the scalar field takes the standard
form:
\ba
\label{ConserEq}
\delta \ddot \varphi+k^2 \delta \varphi+2 a H \delta \dot
\varphi+\dot \varphi_0 \(\dot \Phi+3 \dot \Psi \)+4 a H \dot
\varphi_0 \Phi=0\,.
\ea
The presence of anisotropy \eqref{Psi Phi} will thus modify slightly
this relation by adding an additional term proportional to $r$:
\ba
\label{ConserEqAnisotropy}
\delta \ddot \varphi+k^2 \delta \varphi+2 a H \delta \dot
\varphi+4\dot \varphi_0 \,  \dot \Psi+4 a H \dot
\varphi_0 \Psi=r\, \frac\kf\ell \frac{\dot \varphi_0}{3 a^2H^2}\left[
\delta \ddot \varphi + 2 a H \delta \dot \varphi\right]\,.
\ea
However, as seen previously, at short wavelengths $r\sim -\ell H /
\sqrt{k\tau}$, and this additional contribution on the right hand
side will therefore be negligible. The equation of motion of the
scalar field will thus be unaffected by the anisotropy at short
wavelengths. Since the at long wavelengths the anisotropy vanishes,
we therefore conclude that the equation of motion for the scalar
field will be unaffected by the presence of the Weyl tensor, apart
in the very short transition regime between the short and the long
wavelength regime. But since this transition occurs very rapidly,
one can to a very good approximation match both regimes directly.

%
%


\vspace{20pt}
\begin{thebibliography}{99}



\bibitem{Randall:1999vf}
  L.~Randall and R.~Sundrum,
  ``An alternative to compactification,''
  Phys.\ Rev.\ Lett.\  {\bf 83}, 4690 (1999)
  [arXiv:hep-th/9906064].

%

\bibitem{lowenergy}
  S.~Kanno and J.~Soda,
  ``Brane world effective action at low energies and AdS/CFT,''
  Phys.\ Rev.\ D {\bf 66}, 043526 (2002)
  [arXiv:hep-th/0205188],
%
  T.~Shiromizu and K.~Koyama,
  ``Low energy effective theory for two brane systems: Covariant curvature
  formulation,''
  Phys.\ Rev.\ D {\bf 67}, 084022 (2003)
  [arXiv:hep-th/0210066],
%
 G.~A.~Palma and A.~C.~Davis,
  ``Low energy branes, effective theory and cosmology,''
  Phys.\ Rev.\ D {\bf 70}, 064021 (2004)
  [arXiv:hep-th/0406091].



\bibitem{Barvinsky:2002kh}
  A.~O.~Barvinsky, A.~Y.~Kamenshchik, A.~Rathke and C.~Kiefer,
  ``Braneworld effective action: An alternative to Kaluza-Klein reduction,''
  Phys.\ Rev.\ D {\bf 67}, 023513 (2003)
  [arXiv:hep-th/0206188].

\bibitem{deRham:2004yt}
  C.~de Rham,
  ``Beyond the low energy approximation in braneworld cosmology,''
  Phys.\ Rev.\ D {\bf 71}, 024015 (2005)
  [arXiv:hep-th/0411021].


\bibitem{Cline:2006hu}
  J.~M.~Cline,
  ``String cosmology,''
  arXiv:hep-th/0612129.


%


\bibitem{braneinflation}
  R.~Maartens, D.~Wands, B.~A.~Bassett and I.~Heard,
  Phys.\ Rev.\ D {\bf 66}, 010001 (2002)
  arXiv:hep-ph/9912464,
%
Y.~Shtanov,
``On brane-world cosmology,''
[arXiv:hep-th/0005193],
%
D.~Langlois, R.~Maartens and D.~Wands,
``Gravitational waves from inflation on the brane,''
Phys.\ Lett.\ B {\bf 489}, 259 (2000) [arXiv:hep-th/0006007],
%
E.~Copeland, A.~Liddle and J.~Lidsey,
 ``Steep inflation: ending braneworld inflation by gravitational particle production''
Phys.\ Rev.\ D {\bf 64}, 023509 (2001) [arXiv:astro-ph/0006421],
%
A.~Liddle and A.~Taylor,
``Inflaton potential reconstruction in the braneworld scenario,''
Phys.\ Rev.\ D {\bf 65}, 041301 (2002) [arXiv:astro-ph/0109412],
%
M.~Sami, N.~Dadhich and T.~Shiromizu,
 ``Steep inflation followed by Born-Infeld Reheating
Phys.\ Lett.\ B {\bf 568}, 118 (2003) [arXiv:hep-th/0304187],
%
  K.~Koyama, D.~Langlois, R.~Maartens and D.~Wands,
  ``Scalar perturbations from brane-world inflation,''
  Phys.\ Rev.\ D {\bf 66}, 010001 (2002)
  arXiv:hep-th/0408222.

%


\bibitem{Garriga:1999yh}
  J.~Garriga and T.~Tanaka,
  ``Gravity in the brane-world,''
  Phys.\ Rev.\ Lett.\  {\bf 84}, 2778 (2000)
  [arXiv:hep-th/9911055].

%

\bibitem{Shiromizu:1999wj}
  T.~Shiromizu, K.~I.~Maeda and M.~Sasaki,
  ``The Einstein equations on the 3-brane world,''
  Phys.\ Rev.\ D {\bf 62}, 024012 (2000)
  [arXiv:gr-qc/9910076].



\bibitem{solution}
  N.~Kaloper,
  ``Bent domain walls as braneworlds,''
  Phys.\ Rev.\ D {\bf 60}, 123506 (1999)
  [arXiv:hep-th/9905210],
  \bibitem{solution2}
  O.~DeWolfe, D.~Z.~Freedman, S.~S.~Gubser and A.~Karch,
  ``Modeling the fifth dimension with scalars and gravity,''
  Phys.\ Rev.\ D {\bf 62}, 046008 (2000)
  [arXiv:hep-th/9909134].



\bibitem{Israel:1966rt}
W.~Israel,
Nuovo Cim.\ B {\bf 44S10}, 1 (1966)
[Erratum-ibid.\ B {\bf 48} (1967\ NUCIA,B44,1.1966) 463].


%

\bibitem{BraneFriedEq}
P.~Binetruy, C.~Deffayet and D.~Langlois,
``Non-conventional cosmology from a brane-universe,''
Nucl.\ Phys.\ B {\bf 565}, 269 (2000) [arXiv:hep-th/9905012],
%
P.~Binetruy, C.~Deffayet, U.~Ellwanger and D.~Langlois,
``Brane cosmological evolution in a bulk with cosmological constant,''
Phys.\ Lett.\ B {\bf 477}, 285 (2000) [arXiv:hep-th/9910219],
%
S.~Mukohyama,
 ``Brane-world solutions, standard cosmology, and dark radiation,''
Phys.\ Lett.\ B {\bf 473}, 241 (2000) [arXiv:hep-th/9911165],
%
M.~Sasaki, T.~Shiromizu and K.~Maeda,
 ``Gravity, Stability and Energy Conservation on the Randall-Sundrum Brane-World''
Phys.\ Rev.\ D  {\bf 62}, 024008 (2000) [arXiv:hep-th/9912233],
%
A.~Mazumdar and J.~Wang,
``A note on brane inflation,''
Phys.\ Lett.\ B {\bf 490}, 251 (2000) [arXiv:gr-qc/0004030],
%
P.~Binetruy, C.~Deffayet and D.~Langlois,
``The radion in brane cosmology,''
Nucl.\ Phys.\ B {\bf 615}, 219 (2001) [arXiv:hep-th/0101234],
%
D.~Langlois,
``Brane cosmology: An introduction,''
Prog.\ Theor.\ Phys.\ Suppl.\  {\bf 148}, 181 (2003)
[arXiv:hep-th/0209261],
%
D.~Langlois,
``Cosmology in a brane-universe,''
Astrophys.\ Space Sci.\  {\bf 283}, 469 (2003)
[arXiv:astro-ph/0301022],
%
R.~Maartens,
``Brane-world gravity,''
Living Rev.\ Rel.\  {\bf 7}, 7 (2004) [arXiv:gr-qc/0312059].


\bibitem{CST}
    T.~Battefeld, C.~de Rham and S.~Watson, in preparation.



\bibitem{BD}
N.~Birrell and P.~Davies,  ``Quantum Fields in Curved Space''
Cambridge University Press (1982).


%
\bibitem{Inflation}
  V.~F.~Mukhanov, H.~A.~Feldman and R.~H.~Brandenberger,
  ``Theory Of Cosmological Perturbations. Part 1. Classical Perturbations. Part
  2. Quantum Theory Of Perturbations. Part 3. Extensions,''
  Phys.\ Rept.\  {\bf 215}, 203 (1992),
%
A.~Liddle and D.~Lyth,  ``Cosmological Inflation and Large-Scale
Structure'' Cambridge University Press (2000),
%
 D.~Langlois,
  ``Inflation, quantum fluctuations and cosmological perturbations,''
  arXiv:hep-th/0405053,
  B.~A.~Bassett, S.~Tsujikawa and D.~Wands,
  ``Inflation dynamics and reheating,''
  Rev.\ Mod.\ Phys.\  {\bf 78}, 537 (2006)
  [arXiv:astro-ph/0507632].




\bibitem{Koyama:2007as}
  K.~Koyama, A.~Mennim, V.~A.~Rubakov, D.~Wands and T.~Hiramatsu,
  ``Primordial perturbations from slow-roll inflation on a brane,''
  arXiv:hep-th/0701241.

\end{thebibliography}
\end{document}